\documentclass{wseas}

\usepackage{wseas}
\usepackage[dvips]{epsfig}

\title{ Class numbers in the imaginary quadratic f\mbox{}ield\\
and the $1/f$ noise of an electron gas }

\author{Michel PLANAT}

\begin{document}

  \department{Laboratoire de Physique et Métrologie des Oscillateurs du CNRS }
  \university{associ\'{e} \`{a} l'Universit\'{e} de Franche-Comt\'{e}}
  \address{32 Avenue de l'Observatoire, 25044 Besan\c{c}on Cedex}
  \country{FRANCE}
  \internet{planat@lpmo.edu}

  \summary{Partition functions $Z(x)$ of statistical mechanics are generally
approximated by integrals. The approximation fails in small
cavities or at very low temperature, when the ratio $x$ between
the energy quantum and thermal energy is larger or equal to unity.
In addition, the exact calculation, which is based on number
theoretical concepts, shows excess low frequency noise in
thermodynamical quantities, that the continuous approximation
fails to predict. It is f\mbox{}irst shown that Riemann zeta
function is essentially the Mellin transform of the partition
function $Z(x)$ of the non degenerate (one dimensional) perfect
gas. Inverting the transform leads to the conventional  perfect
gas law. The degeneracy has two aspects. One is related to the
wave nature of particles: this is accounted for from quantum
statistics, when the de Broglie wavelength exceeds the mean
distance between particles. We emphasize here the second aspect
which is related to the degeneracy of energy levels. It is given
by the number of solutions $r_3(p)$ of the three squares
diophantine equation, a highly discontinuous arithmetical
function. In the conventional approach the density of states is
proportional to the square root of energy, that is $r_3(p)\simeq 2
\pi p^{1/2}$. We found that the exact density of states relates to
the class number in the quadratic f\mbox{}ield $Q(\sqrt{-p})$. One
f\mbox{}inds $1/f$ noise around the mean value.}

  \keywords{Electronic circuits, number theory, quantum statistical physics, $1/f$ noise}

  \maketitle

\section{Introduction to Noise \\ in Electrical Circuits}
\label{Electrical} \vspace*{-0.5pt} \noindent

Noise in electrical circuits is found in many forms, some of them
have been well explained: thermal noise, shot noise, partition
noise, burst noise... and one which is still subject to much
debate due to its universality and the lack of a general accepted
model: 1/f noise.

Let us f\mbox{}irst review brief\mbox{}ly our understanding of
electrical thermal noise. Due to thermal agitation free electrons
in a metallic conductor are moving around continuously causing
collisions with the atoms and a continuous exchange of energy
between the modes. This was f\mbox{}irst investigated
experimentally by Johnson \cite{Joh.28} and theoretically
explained by Nyquist \cite{Nyq.28}. The noise in any circuit kept
at uniform temperature $T$ can be described by a noise
voltage$(\overline{v^2})^{1/2}$  in series with a resistor R of
the circuit such that for a small frequency interval $df$
\begin{equation}
\overline{v^2}~=~4~kRTp(f)~df, \label{thermal}
\end{equation}
where $p(f)=\frac{hf}{kT}(e^{hf/kT}-1)^{-1}$ is the Planck factor,
$k=1.38~10^{-23}$ \rm{J/K} is Boltzmann's constant and
$h=6.62~10^{-34}$ \rm{J.s} is Planck's constant. For room
temperature and not too high frequency $hf/kT \ll 1$, so that
$p(f)$ is equal to unity. These equations are  verif\mbox{}ied
from experiments. For example with a resistor of $R=1~K\Omega$, at
room temperature $T=300~K$, if one uses a f\mbox{}ilter of
bandwith $df=1~MHz$, the expected voltage f\mbox{}luctuation is of
the order $4~\mu V$ easily detectable on an oscilloscope.

Our second example is shot noise in a diode. Such a noise is
produced by random emission of electrons from the cathode to the
anode. Schottky theorem  \cite{Sch.18} states that the mean
current $I$ undergoes a f\mbox{}luctuation
\begin{equation}
\overline{i^2}~=~2~eI~df, \label{shot}
\end{equation}
where $e=1.6~10^{-19}$ C is the charge of one electron. Using a
typical value $I=1~mA$ and the same bandwith as above the expected
current f\mbox{}luctuation is on the order $18~nA$.

Our third exemple is partition noise which occurs any time a
current is distributed randomly between two electrodes. Its net
effect is an extra multiplicative factor in the relation for the
shot noise current \cite{Van.70}. A modern variant of shot noise
and partition noise is quantum  partition noise. Eq. (\ref{shot})
was used as an extra proof that the electrical charge is
indivisible and the measurement of current f\mbox{}luctuations
allowed the determination of the unit $e$. Recently it has been
argued that low dimensional systems like quantum wires or quantum
dots may produce a fractional charge $e\frac{p}{q}$ ($p$ and $q$
integers) associated to a quasi particle tunneling state. The
corresponding fractional shot noise current has been observed
\cite {Sam.97}.

In most small size transistors a noise of the low frequency type
is observed: burst noise also called random telegraph signal (or
RTS noise). The noise typically consists of random
discrete-switching events in the current f\mbox{}lowing through a
transistor. It was f\mbox{}irst seen in reverse-biased p-n
junctions and bipolar transistors \cite{Buc.83}. It was then
related to random single electron capture events into localized
defects of very small transistors such as MOSFET's \cite{Kir.89}.
Assuming that to the probability of an electron making a
transition from one state of amplitude $0$ to another state of
amplitude I there is associated a switching time $\tau_1$. The
power spectral density is found to be of the Lorentzian type
\begin{equation}
S_I(f)=\frac{2~I^2 \tau_1}{4+(2\pi f \tau_1)^2}, \label{RTS}
\end{equation}
that is a f\mbox{}lat spectrum up to the turn over frequency and a
$1/f^2$ dependance above.

We now turn to $1/f$ or f\mbox{}licker noise. Examples above refer
to a constant power spectral density $S_V(f)=4kRT$ for thermal
noise and $S_I(f)=2eI$ for shot noise. F\mbox{}licker effect is
the large ammount of noise generated in any solid state device
(vacuum tube, diode, transistor...) at low frequencies which was
discovered by Johnson \cite{Joh.28}. It is generally described
phenomenologically as
\begin{equation}
S_V(f)~=~K\frac{V^a}{f^b}, \label{1/f noiseshot}
\end{equation}
with $a$ close to $2$ and $b$ close to $1$. By integrating between
the low frequency cut-off $1/\tau$ and the high frequency cut-off
$f_c$, we get
\begin{equation}
\overline{v^2}~=~\int_{1/\tau} ^{f_c} S_V(f~)~df \sim K V^2
ln(f_c\tau), \label{power}
\end{equation}
that is the variability in f\mbox{}licker noise depends
logarithmically on the integration time $\tau$, as was observed
experimentally \cite{Kle.88}. Until now $1/f$ noise in solid state
systems eluded attempts to f\mbox{}ind its universel origin.
Fundamental experiments\cite{Vos.76} found f\mbox{}licker noise
associated to thermal noise both in metals and semiconductors
without a clear explanation on its microscopic origin.

From many experiments it was found that the noise intensity is
inversely proportional  to the total number of carriers $N_0$ in
the sample that is $K\sim\gamma/N_0$ with $\gamma$ the Hooge
factor \cite{Hoo.69}.  Values of $\gamma$ in the range $10^{-3}$
to $10^{-8}$ have been found. They were generally attributed to
different scattering mechanisms, by the lattice or the impurities,
leading to mobility f\mbox{}luctuations of the electrons. This
claims for a non linear origin of the $1/f$ noise. F\mbox{}ine
structures revealing the interaction of electrons with bulk and
surface phonons were observed in different solid-state physical
systems by Mihaila \cite{Mih.99}. On the theoretical side a
quantum electrodynamical theory was developed by P. Handel
\cite{Han.80} based on infrared divergent coupling of the
electrons to the electromagnetic f\mbox{}ield in the scattering
process. The basic result for the $\gamma$ parameter is as follows
\begin{equation}
\gamma = \frac{4\alpha}{3\pi}~\frac{\overline{(\delta w)^2}}{c^2},
\label{Handel}
\end{equation}
where $\delta w$ is the change of the velocity of the accelerated
charge in the collision processes governing the carrier mobility
and the carrier diffusion constant, $c= 3.10^8$~\rm{m/s} is the
velocity of light, $\alpha=\frac{e^2}{2~hc\epsilon_0} \simeq
1/137$ is the f\mbox{}ine structure constant and $\epsilon_0 =
8.85~10^{-12}~F/m$ is the permittivity of free space. Thirty years
after its introduction the theory is still under debate
\cite{Kiss},\cite{Weiss}.

In large devices such as MOSFET's $1/f$ noise is often explained
as arising from a summation of RTS's due to the defects in the
insulator whenever a large range of time constants $\tau_1$ is
involved.

$1/f$ noise is also observed in oscillators. One way to
characterize it is from the Allan variance
\begin{equation}
\sigma_y^2(\tau)=\frac{1}{2}{\overline{ \delta y(\tau)^2}},
\label{Allan}
\end{equation}
where $\delta y(\tau)=y_{i+1}(\tau)-y_i(\tau)$ is the deviation
between two consecutive samples of index $i$ and $i+1$ counted
over an averaging time $\tau$ and $y_i=f_i/f_0$ refers to the
ratio between the instantaneous frequency and the mean frequency
of the oscillator. It can be shown \cite{All.87} that $1/f$ noise
of the power spectral density $S_y(f)=C/f$, with C a constant
related to the physical set-up, is associated to a constant value
of the  Allan deviation $\sigma_y(\tau)\sim2~ln2~C$. This
so-called f\mbox{}licker f\mbox{}loor is the limit to the
stability of oscillators. Values in the range $10^{-12}$ for
quartz oscillators to $10^{-15}$  for Hydrogen maser based
oscillators have been obtained \cite{Audoin98}.

We recently discovered a possible relationship between $1/f$
frequency noise in oscillator measurements and prime number theory
\cite{Planat1}, \cite{PRE02}. In this paper we pursue this quest
by relating exact statistical mechanics of electrons of mass $m$
in a box of size $L$ to the quadratic f\mbox{}ield $Q(\sqrt{-p})$
of negative discriminant, with $p$ the number of energy quanta in
units of $\delta E=h^2/8mL^2$. We f\mbox{}ind numerically a $1/f$
noise of the fractional density of states about the average
classical value. Sec. 2 shows that analytical number theory (and
the Riemann zeta function) underlies the statistical mechanics of
the perfect non degenerate gas. Sec. 3 shows that the degeneracy
of energy levels needs the extended frame of algebraic number
theory (and the f\mbox{}ield of quadratic forms). To let the text
accessible to a large audience of readers we do not enter into the
mathematical details that can be found elsewhere
\cite{Edwards},\cite{Grosswald}.

\section{The Non Degenerate Perfect Gas}
\label{nondegenerate} \vspace*{-0.5pt} \noindent
\subsection{The classical approach \\ to the perfect gas}
In statistical thermodynamics the partition function allows to
establish all thermodynamical properties \cite{Kestin}. It is
called generating function in the context of number theory. For a
perfect classical gas the partition function per degree of freedom
and per particle is
\begin{equation}
Z_0=\sum_{n=1}^{n_{\rm{max}}} \exp{(\frac{-E_n}{kT})}.
\end{equation}
The energy levels are obtained by solving Schrödinger equation
with free boundary conditions
\begin{eqnarray}
&&\frac{d^2\psi}{dx^2}+\frac{2mE}{\hbar^2}\psi =0,\nonumber \\ &&
\psi=0~~\rm{at}~x=0~\rm{and}~x=L.
\end{eqnarray}
The solution is
\begin{equation}
\psi=(\frac{2}{L})^{1/2}\sin(\frac{n\pi x}{L}),
\end{equation}
with $E_n=\delta E~ n^2$, $(n=1,2,\cdots)$ so that the partition
function becomes
\begin{equation}
Z_0=\sum_{n=1}^{n_{\rm{max}}} \exp{(-\pi n^2 x) },
\end{equation}
with $x=\frac{\delta E}{kT}$, $\delta E=\frac{h^2}{8mL^2}$.Using
the f\mbox{}irst order Mac-Laurin expansion and if $x$ is small
\begin{equation}
Z_0(x)\simeq \frac{1}{2\sqrt x}\rm{erf}(n_{\rm{max}}\sqrt {\pi x})
-\frac{1}{2}\simeq \frac{1}{2\sqrt x}. \label{MacL}
\end{equation}
%

For the three dimensional gas of $N$ particles we need the
indistinguishability parameter $N!$ so that

\begin{equation}
Z =\frac{Z_0^{3N}}{N!}.
\end{equation}

The gas pressure is def\mbox{}ined from
\begin{equation}
P(V,T)=3NkT(\frac{\partial \ln Z}{\partial
V})_{T,N}~~~~\rm{with}~V=L^3,
\end{equation}

The law of perfect gas follows
\begin{equation}
PV=RT~~~~\rm{with}~R=kN_A,
\end{equation}
where $N_A$ is the Avogadro constant.

Others useful state equations are derived such as the mean energy
per mole of the gas
\begin{equation}
E=-\frac{\partial \ln Z}{\partial
\beta}=\frac{3}{2}RT~~~~\rm{with}~\beta=\frac{1}{kT},
\end{equation}
and the free energy $F$ and the entropy $S$ as
\begin{equation}
S=-\frac{\partial F}{\partial T},~~~~F=-kT \ln Z.
\end{equation}
The Sackur-Tetrode equation follows
\begin{eqnarray}
&S=S(E,V,N)\nonumber \\&=Nk\left[ \frac{5}{2}+\ln(\frac
{V}{N})+\frac{3}{2}\ln(\frac{m}{3\pi\hbar^2}\frac{E}{N})\right].
\end{eqnarray}
Thanks to the inclusion of $N!$ term in the def\mbox{}inition of
partition function the entropy equation satisf\mbox{}ies the
property of extensivity.

\subsection{The classical law of the perfect gas \\ and the Riemann zeta function}

The partition function $Z_0$ appears in Riemann's pioneering work
on prime number theory. It starts from the integral representation
of gamma function
\begin{equation}
\Gamma (s)=\int_0^{\infty}\exp(-x) x^{s-1} dx~~~~\rm{with}~
\Re(s)>0, \label{gamm}
\end{equation}
which generalizes the factorial function into the complex plane
$s$. In particular $\Gamma(s+1)=s\Gamma(s)$ and as result
$\Gamma(n+1)=n!$ if $n$ is an integer.

Riemann's zeta function is def\mbox{}ined as
\begin{equation}
\zeta(s)=\sum_{n=1}^{\infty}\frac{1}{n^s}~~~~\rm{with}~\Re(s)>1.
\end{equation}
Substituting $\pi n^2 x$ to $x$ in the integral (\ref{gamm}) and
summing for all integers $n$ one gets
\begin{equation}
\pi^{-s} \Gamma(s) \zeta(2s)=\int_0^{\infty} Z_0(x) x^{s-1} dx,
\end{equation}
which is the desired relation between $\zeta(s)$ and the Mellin
transform of $Z_0(x)$. Here we assumed $n_{\rm{max}} \rightarrow
\infty$ in the def\mbox{}inition of the partition function. The
expression
\begin{equation}
\xi(s)=\pi^{-s/2} \Gamma(s/2)\zeta(s),
\end{equation}
is the so-called completed Riemann zeta function. This is because
$\xi(s)$ admits an analytic continuation to the whole complex
plane $s$ except at the simple poles $s=0$ and $s=1$ of residus
$-1$ and $1$ respectively. It satisf\mbox{}ies the functional
relation $\xi(s)=\xi(s-1)$.

The approximation of $Z_0(x)$ may be obtained by taking the
inverse Mellin transform and by applying the residue theorem to
the two poles of $\xi(2s)$ at $s=0$ and $s=1/2$ so that
\begin{equation}
Z_0(x) \simeq \frac{1}{2 \sqrt x}-\frac{1}{2},
\end{equation}
which is similar to the previous Mac-Laurin approximation
(\ref{MacL}). The step from the discrete to the continuous is
condensed into the contribution of poles of the Riemann zeta
function.

The remaining poles at $s=-m$ ($m$ integer $\ge 1$) of $\Gamma(s)$
are absorbed by the trivial zeros at $s=-l$ ($l$ a positive
integer) of $\zeta(2s)$. The unproven Riemann hypothesis
\cite{Planat1} is that all remaining zeros are located on the
critical line $s=1/2$. The modulating action of the zeros is
expected to be signif\mbox{}icant within a box at the nanoscopic
scale $x=\frac{\delta E}{kT}\ge 1$.

\section{The Degenerate Perfect Gas}
\vspace*{-0.5pt} \noindent
\subsection{F\mbox{}irst form of quantum degeneracy \\ due to wave-particle aspects}

In the quantum gas, the f\mbox{}irst form of degeneracy is due to
the indiscernability of (wave)particles of size
$\lambda_{\rm{th}}$ (the de Broglie wavelength) larger than the
typical spacing $d$. The degeneracy parameter is \cite{Kestin}
\begin{equation}
\frac{\lambda_{\rm{th}}}{d}=\frac{h/(mv)}{V^{1/3}},~~\rm{with}~v=(\frac{2
\pi k T}{m})^{1/2}.
\end{equation}
(For conduction electrons in silver at room temperature the ratio
is about $40$).

The partition function for electrons is taken to satisfy
Fermi-Dirac statistics with chemical potential $\alpha$
\begin{equation}
\ln Z=\sum_s \ln \left[ 1+ \exp(-\alpha -\beta E_s) \right] +N
\alpha.
\end{equation}
In the conventional approach of statistical mechanics, the
discretness of quantum states is neglected and one goes from the
discrete to the continuous by taking the number of states
$\Omega_0(E)$ as $\frac{1}{8}$, times the volume of the Fermi
sphere of radius $R_F$, times $2$ (for the two spin states of the
electron). The density of states $\omega_0(E)$ per unit volume
$V=L^3$ follows from
\begin{equation}
 \omega_0(E)=\frac{1}{V}\frac{d \Omega_0(E)}{dE}=2 \pi
 (\frac{2m}{h^2})^{3/2} E^{1/2}.
\label{equation2}
\end{equation}
%
\subsection{Second quantum aspect \\ due to the degeneracy of energy levels}
Here the number of possible quantum states is calculated exactly
from number theory.
\begin{equation}
\Omega(E)=\sum_{n_1^2+n_2^2+n_3^2 \le R_F^2}
1~~\rm{with}~~R_F^2=\frac{E_F}{\delta E}, \label{equation1}
\end{equation}
where $n_1$, $n_2$ and $n_3$ are positive integers, $E_F$ is the
Fermi energy and $\delta E$ is the energy quantum as above. In
quantum dots the energy is quantized in units of $\delta E$: this
may be observed when the Coulomb blockade energy $E_B=\frac{e^2}{
C}$ is such that $E_B \simeq \delta E \gg kT$, that is at very low
temperatures or for small samples with large capacity $C$.
 The partition function is
\begin{equation}
Z=\sum_{p=1}^{p_{\rm{max}}} g_p\exp
(-\frac{E_p}{kT})~~\rm{with}~E_p=\delta E ~ p,
\end{equation}
where the degeneracy parameter $g_p$ is the number of states of
energy $E_p$.
\begin{figure}[htbp]
\rotatebox{-90}{\resizebox{5cm}{!}{\includegraphics{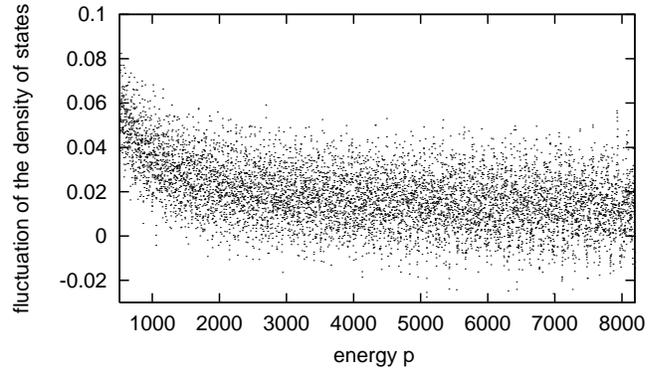}}}
\caption{Relative f\mbox{}luctuations $\epsilon (p)$ of the
density of states versus energy $p$ (We took $\delta p=100$).}
\end{figure}

The number of quantum states $g_p$ also equals $\frac{1}{8}$,
times 2, times the number $r_3(p)$ of solutions of the three
squares diophantine equation
\begin{equation}
 n_1^2+n_2^2+n_3^2=p ~~~\rm{with}~~ p\le R_F^2,
\label{equation3}
\end{equation}
\begin{figure}[htbp]
\rotatebox{-90}{\resizebox{5cm}{!}{\includegraphics{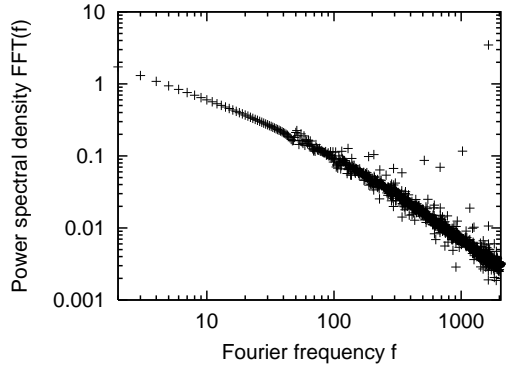}}}
\caption{Power spectral density of the f\mbox{}luctuations}
\end{figure}
and $n_1,~n_2,~n_3~,p$ integers. This is a central problem of
number theory. Gauss proved that for $p>4$ and squarefree
\begin{equation}
r_3(p)= \left\{\begin{array}{ll} 24h(-p) &~~\mbox{for}~ p\equiv
3~(mod~8),\\ 12  h (-4p)&~~ \mbox{for}~p \equiv 1,2,5~\rm{or}~6
~(mod~8),
\\ 0&~~ \mbox{for}~ p\equiv 7~ (mod~8),
\end{array}\right.
\label{equation4}
\end{equation}
where $h(-p)$ is the class number in the f\mbox{}ield
$Q(\sqrt{-p})$ (that is the number of primitive quadratic forms
with given negative discriminant -p). The general case is
$r_3(p)=\sum _{d^2|p}R_3(\frac{p}{d^2})$, where $R_3(p)$ holds for
square free values of $p$. More explicit solutions can be found
\cite{Grosswald}. They are available in the Mathematica package
\cite{Wolfram}.

The class number $h(-p)$, and thus $r_3(p)$ is a highly
discontinuous function. We def\mbox{}ined the (exact) density of
states as

\begin{eqnarray}
&\omega(p)=\frac{1}{\delta p}\sum_{n=p}^{n=p+\delta p}
r_3(n)\nonumber \\ &\simeq \omega_0(p)=2\pi
p^{1/2}~~\rm{with}~~~\delta p \ll p. \label{equation5}
\end{eqnarray}

Relative f\mbox{}luctuations of the density of states around the
mean value $\omega_0(p)$ were calculated as
$\epsilon(p)=\frac{\omega(p)-\omega_0(p)}{\omega_0(p)}$ (see
F\mbox{}ig. 1).
As a result the power spectral density of low frequency
f\mbox{}luctuations versus Fourier frequency $f$ was found as $FFT
(f)\sim \frac{A~ \delta p}{f^{\alpha}}$, with $\alpha$ close to
$1$ and $A\sim 0.1$, that is a $1/f$ spectrum as found in
F\mbox{}ig. 2.

Physically one can take $\delta p=[\frac{kT}{\delta E}]$, with
$[~]$ the integer part and $\delta E$ the granularity parameter,
since $kT$ is the width of the Fermi surface at f\mbox{}inite
temperature T. One can conclude that the $1/f$ spectrum arising in
the arithmetical approach connects to the different regimes one
gets for the electron gas in quantum dots, versus the typical size
$L$ of the dot, the temperature $T$ and the effective mass $m$.

Due to the low frequency noise in the density of states all
thermodynamical quantities are perturbed, as well as the
conductivity of electronic devices. This will be developed in a
future paper.

\section{Concluding Remarks}
\vspace*{-0.5pt} \noindent Thermodynamical quantities attached to
the quantum description of an electronic gas in a cubic box are
not smooth versus energy or time, contrary to the classical
viewpoint: this is because the summatory function of an
arithmetical function such as $r_3(n)$ is discontinuous. The
fractional deviation with respect to the asymptotic average very
often leads to a low frequency power spectrum of the fractal type
$1/f^b$, $b$ close to $1$. Number theory thus is a good candidate
to explain many low frequency noises encountered in solid state
physics and engineering as well as in others f\mbox{}ields.

\section*{Acknowledgements}
I acknowledge Rémi Barrère for his help in the programming of
arithmetical functions used in this paper.

\end{document}